\documentclass[copyright]{eptcs}
\providecommand{\event}{PLACES 2012}

\usepackage{breakurl}

\usepackage{color}
\usepackage{amsthm}
\usepackage{amsmath}
\usepackage{amssymb}
\usepackage{listings}
\usepackage{enumerate}
\usepackage{textcomp}

\newtheorem{theorem}{Theorem}[section]
\newtheorem{definition}{Definition}[section]

\definecolor{color:keyword}{rgb}{0.53,0.05,0.05}
\definecolor{color:comment}{rgb}{0.25,0.37,0.75}
\definecolor{color:string}{rgb}{0.87,0.0,0.0}

\lstdefinelanguage{Jolie}{
morekeywords={
	OneWay,RequestResponse,
	main,define,inputPort,outputPort,init,execution,include,
	cset,interface,type,throws,global,constants,for,foreach,while,new
},
sensitive=true,
morecomment=[l]{//},
morecomment=[s]{/*}{*/},
morestring=[b]",
otherkeywords={;,|,:}
}

\lstdefinelanguage{Chorus}{
morekeywords={
	rec,if,else,starts,new,init,via
},
sensitive=true,
morecomment=[l]{//},
morecomment=[s]{/*}{*/},
morestring=[b]",
otherkeywords={;,|,:,->}
}

\newcommand{\lstcode}[1]{\lstinline~#1~}

\newcommand{\lblread}[1]{\textsf{read} \ v}

\newcommand{\nil}{\boldsymbol 0}

\newcommand{\Dids}[2]{\lfloor^{\textsc{#1}}|{\textsc{{\tiny #2}}}\rceil}

\newcommand{\function}[2]{\mathsf{#1}(#2)}

\newcommand{\rec}[1]{\textsf{rec}\, #1\, \textsf{in}\,}
\newcommand{\REC}[2]{\textsf{rec}\, \mathbf{#1}\, \mathsf{in}\, #2}
\newcommand{\RECVAR}[1]{\mathbf{#1}}

\newcommand{\arr}[1]{\mathrel{\stackrel{{\;\;#1\;\;}}{\mbox{\rightarrowfill}}}}

\newcommand{\trans}[2]{#1 \, \arr{} \, #2}

\newcommand{\transMulti}[2]{#1 \, \arr{}^{*} \, #2}

\newcommand{\res}[1]{\boldsymbol( \boldsymbol\nu #1 \boldsymbol)\:}

\newcommand{\typeint}{\mathbf{int}}

\newcommand{\typebool}{\mathbf{bool}}
\newcommand{\typestring}{\mathbf{string}}

\newcommand{\typefile}{\mathbf{file}}

\newcommand{\code}[1]{\texttt{#1}}
\newcommand{\start}[4]{#1\;\mathbf{{start}}\;#2#3(#4)}
\newcommand{\genericstart} {\start
  {\tau_1[\role p_1],\ldots,\tau_n[\role p_n]}{}ak}

\newcommand{\interact}[3]{#1\;\code{->}\;#2:#3}
\newcommand{\role}[1]{\mathtt{#1}}

\newcommand{\selection}[4]{#1\;\code{->}\;#2:#3[\labels{#4}]}

\newcommand{\genericselection}{\selection{\tau_1[\role p]}{\tau_2[\role
q]}{k}{l}}
\newcommand{\genericselectionPrime}{\selection{\tau_1[\role p]}{\tau_2[\role
q]}{k'}{l}}

\newcommand{\genericinteract}{\interact{\tau_1[\role p].e}{\tau_2[\role
q].x}{k}}
\newcommand{\genericinteractPrime}{\interact{\tau_1[\role p].e}{\tau_2[\role
q].x}{k'}}

\newcommand{\typeinteract}[3]{\interact{\mathtt{#1}}{\mathtt{#2}}{#3}}

\newcommand{\generictypechoice}{\interact{\mathtt{p}}{\mathtt{q}}
  {\{ \labels{l}_i:G_i \}_{i \in I}}}
\newcommand{\generictypeinteract}{\typeinteract{p}{q}{S}}
\newcommand{\inacttype}{\mathsf{end}}

\newcommand{\ifthenelseC}[4]{\textsf{if} \, #2@#1\, \textsf{then}\, #3\,
\textsf{else}\, #4}
\newcommand{\genericifthenelseC}{\ifthenelseC{\tau}{e}{C_1}{C_2}}

\newcommand{\Chor}{\textsf{Chor}}

\newcommand{\NI}{\noindent}
\newcommand{\pp}{\ \boldsymbol{|}\ }
\newcommand{\labels}[1]{\mathit{#1}}
\newcommand{\thr}[1]{\mathsf{#1}}

\newcommand{\tproves}[3]{#1\ \vdash\ #2\ \triangleright\ #3}

\newcounter{ncomm}

\newcommand{\simplify}[2]{{\{\![\, #1\, ]\!\}}^{#2}}
\newcommand{\mesh}[1]{\mathsf{mesh}(#1)}

\newcommand{\paths}[1]{\mathsf{paths}(#1)}

\begin{document}
\title{Merging Multiparty Protocols in Multiparty Choreographies}

\author{Marco Carbone\thanks{Research supported by the Danish Agency
    for Science, Technology and Innovation.}
  \institute{IT University of Copenhagen}
  \institute{Copenhagen, Denmark}
  \email{carbonem@itu.dk}
  \and
  Fabrizio Montesi
  \institute{IT University of Copenhagen}
  \institute{Copenhagen, Denmark}
  \email{fmontesi@itu.dk}
}
\def\titlerunning{Merging Multiparty Protocols in Multiparty Choreographies}
\def\authorrunning{M. Carbone \& F. Montesi}

\maketitle
%
\begin{abstract}
Choreography-based programming is a powerful paradigm for defining 
communication-based systems from a global viewpoint. A choreography
can be checked against multiparty protocol specifications, given as behavioural types,
that may be instantiated indefinitely 
at runtime. Each protocol instance is started with a 
synchronisation among the involved peers. 

We analyse a simple transformation from a choreography with a possibly 
unbounded number of protocol instantiations to a choreography 
instantiating a single protocol, which is the merge of the original 
ones. This gives an effective methodology for obtaining new protocols 
by composing existing ones. Moreover, by removing all synchronisations 
required for starting protocol instances, our transformation reduces 
the number of communications and resources needed to execute a choreography.
\end{abstract}

\section{Introduction}
\label{sec:introduction}
Communication-based programming is a widespread design paradigm
where the entities of a system communicate exclusively by means of
message passing. Communication-based systems are employed in many
areas, from multi-core programming~\cite{mpi} to service-oriented and
cloud computing~\cite{jolie:ecows07,bpel,amqp,paasoa}. In such systems,
entities engage in communication flows
where their input and output operations must follow a specific
order. The structure of each flow is defined by a protocol.

Choreography-based programming is an emerging methodology for defining
communication-based systems in terms of \emph{global
descriptions}.  A global description gives a global view of how
messages are exchanged during execution, in contrast with the
methodologies where the code for each entity is defined separately.
Global descriptions have been studied as
models~\cite{CHY12,BGGLZ06,LGMZ08,HYC08}, as
standards~\cite{WSCDL,BPMN}, and as language
implementations~\cite{HMBCY11,pi4soa,DBLP:conf/tools/NgYH12}.

In~\cite{CM12} we propose a language where both the abstract and the concrete
descriptions of a system are given in global terms.
Programmers can use
\emph{choreographies} for defining the concrete behaviour of a system
and then check them against \emph{protocols}, given as \emph{global types}~\cite{HYC08}. 
The language allows for instantiating
different protocols multiple times, checking that each instantiation
respects the corresponding protocol type. In the sequel, we introduce
a small example for explaining the basic mechanisms of a
choreography and how it integrates with a protocol specification.

Our example implements two protocols, called $G_a$ and $G_b$.  In
$G_a$, a user $\role U$ sends a message to a client application $\role
C$ together with some authentication credentials (encoded as a
string). This is expressed by the global type:
\begin{small} \[G_a\ =\ \typeinteract{U}{C}{\typestring}\] \end{small}
\NI\!\!where $\role U$ and $\role C$ are called the \emph{roles} of
the protocol.  The above behavioural type simply expresses that for
executing protocol $G_a$, whoever plays role $\role U$ needs to send a
message of type $\typestring$ to a party playing role $\role C$.
Protocol $G_b$ is a little bit more complex:
\begin{small}
\[G_b\ =\ 
\typeinteract{C}{F}{\typestring};\quad
\typeinteract{F}{C}{\{ \labels{ok}:
  \typeinteract{C}{F}{\typefile},\quad
  \labels{quit}: \inacttype\}}\] 
\end{small}
\NI \!\!Here, the roles interacting are a client $\role C$ and a file
server $\role F$. In $G_b$, the client sends a
string to the file server $\role F$ (some authentication credentials)
which can reply with either label $\labels {ok}$ or label
$\labels{quit}$. In the first case, the client will send a file to be
stored to the file server, whereas in the other case the protocol will
just terminate.

Although the global types above give a good abstraction of the protocols
that a programmer wishes to use, they give no information on how they
can be combined in an implementation.  A possible
implementation can be given with the choreography $C$ defined below:
  \begin{equation}\label{chor1}
  \begin{small}
  \begin{array}{lllll}
    C \ = \ & 1.\quad &
    \rec {\ X\ } {} \quad &
    \start{\thr {c}[\role{C}],\thr {u}[\role {U}]}{}{a}{k};\\
    & 2.\quad& & \interact{\thr {u}[\role {U}].\function{password}{}}{\thr {c}[\role
      {C}].pwd}{k};\\
    & 3.\quad&&
    \start{\thr {c}[\role{C}],\thr {f}[\role F]}{}{b}{k'};\\
    & 4.\quad&&
    \interact{\thr {c}[\role C].pwd}{\thr {f}[\role
      {F}].y}{k'};\\[.5mm]
    & 5.\quad && \textsf{if }\function{check}{y}@\thr{f}
    \ \ \textsf{then}\ 
    \selection{\thr{f}[\role F]}{\thr {c}[\role C]}{k'}{ok};\\
    & 6.\quad && 
    \phantom{\textsf{if }\function{check}{y}@\thr{f} \ \ \textsf{then}\ }
    \interact{\thr {c}[\role C].file}{\thr {f}[\role
      {F}].z}{k'}\\[.5mm]
    & 7.\quad &&
    \phantom{\textsf{if }\function{check}{y}@\thr{f}}
    \ \  \textsf{else}\ \ \,
    \selection{\thr{f}[\role F]}{\thr {c}[\role C]}{k'}{quit};\ X
  \end{array}
  \end{small}
  \end{equation}
%
We briefly describe the choreography above.
In Line 1, the operation $\start{\thr
  {c}[\role{C}],\thr {u}[\role {U}]}{}{a}{k}$ starts protocol $G_a$
where $\thr c$ and $\thr u$ denote two executing threads willing to
implement protocol $G_a$ playing roles $\role C$ and $\role U$
respectively. The two symbols $a$ and $k$ denote the name of the
protocol $G_a$ and a session identifier $k$ (which functions as a
binder). In Line 2, we can see a communication over session $k$ where
the user $\thr u$, playing role $\role U$, sends the return value of
some internal function $\function{password}$ to $\thr c$. In Line 3,
thread $\thr c$ and another thread $\thr f$ start protocol $G_b$,
similarly to the start of protocol $G_a$. Line 4 contains a new
interaction where $\thr c$ forwards the password $\mathit{pwd}$ to
$\thr f$. In Line 5 we have an if-then-else implementing the abstract
branching given in the protocol description of $G_b$. Note that in the
then-branch $\thr f$ communicates the choice of $\labels{ok}$ to $\thr
c$, whereas it uses label $\labels{quit}$ in the else-branch.

The choreography in (\ref{chor1}) interleaves the two protocols $G_a$
and $G_b$ in a particular order decided by the programmer.  The local
behaviour of each thread (implementation) can be then automatically
generated by means of \emph{EndPoint Projection}~\cite{CM12,CHY12}.
We observe that~(\ref{chor1}), since it executes two protocols, has
two operations for protocol initiation (Lines 1 and 3) in the body of
a recursion. At the endpoint level starts are implemented through
synchronisations between peers~\cite{BCDLDY08,HYC08}, which may be
computationally expensive in a distributed system. Now, we ask:
\begin{quote}\em
  Can we remove the synchronisation points introduced by start operations
  in a choreography? And, what are the consequences?
\end{quote}
In this paper, we analyse a straightforward transformation on
choreographies that cancels out start operations. E.g., the
choreography $C$ in~(\ref{chor1}) could be transformed into:
\begin{small}
  \begin{equation}\label{chor2}
  \begin{array}{lllll}
    1.\quad &
    \start{\thr {c}[\role{C}],\thr {u}[\role {U}],\thr {f}[\role
{F}]&\!\!\!\!\!\!}{}{c}{k};\\
    2.\quad&    \rec {\ X\ } {} \quad  &
    \interact{\thr {u}[\role {U}].\function{password}{}}{\thr {c}[\role
      {C}].pwd}{k};\\
    4.\quad&&
    \interact{\thr {c}[\role C].pwd}{\thr {f}[\role
      {F}].y}{k};\\[.5mm]
    5.\quad && \textsf{if }\function{check}{y}@\thr{f}
    \ \ \textsf{then}\ 
    \selection{\thr{f}[\role F]}{\thr {c}[\role C]}{k}{ok};\\
    6.\quad && 
    \phantom{\textsf{if }\function{check}{y}@\thr{f} \ \ \textsf{then}\ }
    \interact{\thr {c}[\role C].file}{\thr {f}[\role
      {F}].z}{k}\\[.5mm]
    7.\quad &&
    \phantom{\textsf{if }\function{check}{y}@\thr{f}}
    \ \  \textsf{else}\ \ \,
    \selection{\thr{f}[\role F]}{\thr {c}[\role C]}{k}{quit};\ X
  \end{array}
\end{equation}
\end{small}
\NI\!\!Although (\ref{chor2}) has a single start operation, we observe
that it is
semantically related to (\ref{chor1}), since all data communications performed
in (\ref{chor1}) are also performed in (\ref{chor2}) and viceversa.
Moreover, since the single synchronisation point in (\ref{chor2}) 
is no longer under recursion, we conjecture that this
aspect may greatly improve the execution of choreographies in
asynchronous settings.

We also observe that (\ref{chor2}) does no longer implement
the two binary protocols $G_a$ and $G_b$, but it subsumes a new
three-party protocol $G_c$ obtained by composing the former two:
\begin{small}
\[
  G_c \ =\  \REC {\ t\ \ } {\ }\\
  \quad \interact{\role U}{\role C}{\typestring};\ 
  \interact{\role C}{\role {F}}{\typestring};\ 
  \interact{\role F}{\role C}{
    \{
    \labels{ok}: \interact{\role C}{\role F}{\typefile},\quad
    \labels{quit}: \RECVAR t
    \}
  }
\]
\end{small}
\!\!Note that because of the recursive behaviour appearing in $C$ (but not
in $G_a$ and $G_b$), we need to include some recursive behaviour in
the new type (hence the recursion $\mathsf{rec}$ $\mathbf{t}$).
Observe also that $G_c$ is a \emph{multiparty} protocol, i.e. it
considers more than two participants.  Because of this, we believe
that such a transformation could be used for creating new protocols.
In fact, it may happen that such a choreography
represents a pattern that the programmer may want to reuse in other
programs.  Unfortunately, there is no way to reuse such a pattern in a
safe way other than copying and editing the code.  By using the
transformation hinted above, we could abstract the behaviour of a
choreography and make it reusable in other programs.






In the remainder of the paper, we try to lay the foundations of this
idea by giving a formalisation of the concept into a simplified
version of the global calculus with multiparty protocols~\cite{CM12}.

\section{Formalisation and Results}
\label{sec:theory}
\NI\textbf{Calculus, Semantics and Types.}  We formalise our choreographies with a simplification of the Global Calculus
(GC)~\cite{CM12}.
Fig.~\ref{fig:chorsyntax} reports the syntax of GC.
\begin{figure}[h]
  \begin{small}
    \begin{displaymath}
        \begin{array}{r ll ll}
          C\ ::=\ & \ \eta;C & \ \textit{(seq)}
          \\[1mm]
          
          | & \ \genericifthenelseC & \ \textit{(cond)}

          \\[1mm]

          | & \ \rec{\ X\ }{C} & \ \textit{(rec)}

          \\[1mm]

          | & \ X& \
                \textit{(call)}
          
          \\[1mm]

          | & \ \res k C & \ \textit{(res)}
          
          \\[1mm]
          
          | & \ \nil & \ \textit{(inact)}

          \\[3mm]

          \eta\ ::=\ & \ \genericstart & \ \textit{(start)}
          
          \\[1mm]
          
          | & \ \genericinteract & \ \textit{(com)}

          \\[1mm]

          | & \ \genericselection & \ \textit{(sel)}

        \end{array}
    \end{displaymath}
  \end{small}
\caption{Global Calculus, syntax.}
\label{fig:chorsyntax}
\end{figure}

In the Figure, $\tau$ is a thread (running process); $\role{p},\role q, \ldots$
are roles;
$a$ is a public channel; $k$ is a session channel; $x$ is a
placeholder for values; and $\labels{l}$ is a {\em label} for
branching. $e$ denotes a first-order expression on values (integers,
strings, \ldots), whose syntax we leave unspecified.
We read \emph{(seq)} as: do $\eta$ and then proceed as $C$. $\eta$ represents
an interaction between some threads.
Term {\em (start)} denotes the initiation of a multiparty session
(protocol): threads $\tau_i$ wish to start the multiparty session $a$
and tag it with a fresh session channel (identifier) $k$, which is bound in the
choreography continuation. The threads
$\tau_i$ are ordinary threads running in parallel.  The $\role p_i$'s
denote the roles played by the threads in the session.  In-session
communication is denoted by the term \emph{(com)} where thread
$\tau_1$ sends the evaluation of expression $e$ to thread $\tau_2$
which binds it to variable $x$ in the choreography continuation. In term {\em
(sel)}, $\tau_1$
communicates to $\tau_2$ her wish to select branch $\labels{l}$.
In term \emph{(cond)}, thread $\tau$ makes an internal choice between branches
$C_1$ and $C_2$ by evaluating $e$.
\emph{(rec)} and \emph{(call)} model standard recursion.
\emph{(res)}, used only at runtime, allows to bind session channel $k$ in $C$.
We use $\res {k_1,\ldots,k_n}$ as an abbreviation for $\res {k_1} \ldots
\res{k_n}$.
$\nil$ is the empty choreography.
%

The semantics of the global calculus
is a reduction relation $\to$ which just reduces the size of a
choreography wrt prefixing.
Formally, $\to$ is the smallest relation satisfying the rules reported in
Fig.~\ref{fig:chorsemantics}.
\begin{figure}
  \begin{small}
  \begin{displaymath}
    \begin{array}{rlrlrlrlrlrlrlrlrlrlrl}
      \Dids{C}{Start}\
      &
      \genericstart;C \to \res{k}C
      \\[1mm]
      \Dids{C}{Com}\
      &
      \genericinteract;C \to C[v/x] \qquad (e \downarrow v)
      \\[1mm]
      \Dids{C}{Sel}\
      &
      \genericselection;C \to C
      \\[1mm]
    \Dids{C}{If}\
    &
    \genericifthenelseC \ \to\  C_i
    \quad
    (i = 1 \  
    \mbox{if}\  e \downarrow \mbox{true}\, ,\  i = 2\ \mbox{otherwise}
    ) 
    \\[1mm]
    \Dids{C}{Ctx}\
    &
    C \to C'
    \ \Rightarrow\
    \rec{X}{C}
    \ \to \ \rec {X}{C'}
    \\[1mm]
    \Dids{C}{Res}\
    &
    C \to C'
    \quad\Rightarrow\quad
    \res k C \to \res k C'
    \\[1mm]
    \Dids{C}{Struct}\
    &
    C_1 \, \equiv \, C'_1 \quad C'_1 \to C'_2 \quad
    C'_2 \, \equiv \, C_2
    \quad\Rightarrow\quad
    C_1 \to C_2
  \end{array}
  \end{displaymath}
  \end{small}
  \caption{Global Calculus, semantics.}
  \label{fig:chorsemantics}
\end{figure}
In $\Dids{C}{Struct}$, structural congruence
$\equiv$ is standard: it handles alpha-renaming and expansion of recursive
calls.
Given that there are some fresh names
created, the semantics may introduce restriction operators (Rule
$\Dids{C}{Start}$). For instance, the term
\begin{small}
\[
C \quad  = \quad \start{\thr {c}[\role{C}],\thr {u}[\role {U}]}{}{a}{k};\
\interact{\thr {u}[\role {U}].\function{password}{}}{\thr {c}[\role
  {C}].pwd}{k};C'
\]
\end{small}
would have the following reduction chain:
\begin{small}
\[
C \quad \to\quad  \res k\interact{\thr {u}[\role {U}].\function{password}{}}{\thr
  {c}[\role {C}].pwd}{k};C'
\quad\to \quad \res k C'[\function{password}{}/pwd]
\]
\end{small}

In~\cite{CM12}, we develop a type theory for our choreography model exploiting global
types for representing protocols (as we did in the Introduction).
A type system checks that the protocol instances in a choreography follow the given global
types. As an example, we can see that
protocols $G_a$ and $G_b$ are correctly used by the choreography $C$
given in the Introduction.
Fig.~\ref{fig:gt_syntax} reports the syntax for global types.
\begin{figure}[h]
  \begin{small}
    \begin{displaymath}
     \begin{array}{rll}
       G\ ::=\ & \generictypeinteract;G & \ \textit{(com)}
       \\[1mm]
       | & \generictypechoice & \ \textit{(choice)}
       \\[1mm]
       | & \inacttype & \ \textit{(inact)}
       \\[1mm]
       | & \REC{\ \mathbf t\ } G & \ \textit{(rec)}
       \\[1mm]
       | & \mathbf t & \ \textit{(call)}
       \\[3mm]
       S \ ::= \ & \typebool\pp\typeint\pp\typestring\pp\typefile\pp\ldots & \
       \textit{(sort)}
     \end{array}
    \end{displaymath}
  \end{small}
\caption{Global Types, syntax.}
\label{fig:gt_syntax}
\end{figure}

%
\smallbreak
\NI\textbf{Choreography Transformation. } We can now present our transformation for choreographies.  Formally,
we define a function $\simplify{C}{k}$ that transforms a choreography
$C$ into another choreography which implements the same behaviour of
$C$ using a single session $k$.  $\simplify{C}{k}$ is inductively
defined by the following rules.
%
%
Below we assume, without any loss of generality, that all session
channels started in $C$ are different, i.e. there are no two subterms
of the form
$\start{\tau_1[\mathtt{p}_1],\ldots,\tau_n[\mathtt{p}_n]}{}{a}{k}$ in
$C$ with the same $k$.
\begin{displaymath}
\begin{small}
\begin{array}{c}
  \begin{array}{rcl}
    \simplify{\start{\tau_1[\mathtt{p}_1],\ldots,\tau_n[\mathtt{p}_n]}
      {}{a}{k'};C'}{k}
    \ & = & \ 
    \simplify{C'}{k}
    \\[.5mm]
    \simplify{\genericinteractPrime;C}{k}
    \ & = & \ 
    \interact{\tau_1[\tau_1].e}{\tau_2[\tau_2].x}{k};\simplify{C}{k}
    \\[.5mm]
    \simplify{\genericselectionPrime;C}{k}
    \ & = & \ 
    \selection{\tau_1[\tau_1]}{\tau_2[\tau_2]}{k}{l};\simplify{C}{k}
  \end{array}
  \\
  \begin{array}{c}
    \\[-3mm]
    \simplify{\genericifthenelseC}{k}
    \ = \
    \ifthenelseC{\tau}{e}{\simplify{C_1}{k}}{\simplify{C_2}{k}}
    \qquad \simplify{\rec X C}{k}\  =  \
    \rec X \simplify{C}{k}
  \end{array}
\end{array}
\end{small}
\end{displaymath}
%
We briefly comment the rules above.
\emph{(start)} terms are simply removed. In interactions the role of each
thread is annotated with the thread name, in order to maintain the distinction between
roles with the same name played by different threads. All other terms are preserved.

%
%
We can give an example of the transformation by applying it to the
choreography $C$ using protocols $G_a$ and $G_b$ in the Introduction.
We obtain:
\[
\begin{small}
\begin{array}{lllll}
  \simplify{C}{k} = \quad &
  1. & \rec X {}
  \quad \interact{\thr {u}[\role u].\function{password}{}}{\thr
    {c}[\role c].pwd}{k};
  \quad
  \interact{\thr {c}[\role c].pwd}{\thr {f}[\role f].y}{k};\\[.5mm]
  & 2. & \phantom{\rec X {}} \quad \textsf{if }\function{check}{y}@\thr{f}
  \ \ \textsf{then}\ 
  \selection{\thr{f}[\role f]}{\thr {c}[\role c]}{k}{ok};
  \quad 
  \interact{\thr {c}[\role c].file}{\thr {f}[\role f].z}{k}\\[.5mm]
  & 3. &\quad
  \phantom{\rec X {} \textsf{if }\function{check}{y}@\thr{f}}
  \ \  \textsf{else}\,\, \,
  \selection{\thr{f}[\role f]}{\thr {c}[\role c]}{k}{quit};\ X
\end{array}
\end{small}
\]
Observe that the result is the same to the transformation example we
have shown in the Introduction, up to renaming of roles and the first
start operation for starting $k$. We omit how to automatically
generate the latter, since it can be done through a very simple
traversal of the structure of $\simplify{C}{k}$, tracking the roles of
each thread in the interactions.

%
\smallbreak
\NI\textbf{Results.}
Hereby, we present some of the properties enjoyed by our simple
transformation.
%
%
%

Our first result is about the correctness of the behaviour of the transformation result.
Specifically, the transformation does not
introduce any additional behaviour (soundness) and it preserves the original behaviour
up to removal of start terms and renaming of roles (completeness).
\begin{theorem}[Correctness]
Let $C$ be a choreography and $k$ a session channel name. Then,
\begin{itemize}
\item \emph{(Soundness)} $\trans{\simplify{C}{k}}{C'}$ for some $C'$ implies that there
exists $C''$
such that $\trans{C}{C''}$ and $C' = \simplify{C''}{k}$
\item \emph{(Completeness)} $\trans{C}{C'}$ for some $C'$ implies that there exists $C''$
such that
$\transMulti{C'}{C''}$ and $\trans{\simplify{C}{k}}{\simplify{C''}{k}}$
\end{itemize}
\end{theorem}
The result above can also be stated in a stronger form in terms of bisimilarity, using
the labelled semantics reported in~\cite{CM12}. We chose this form for the sake of
brevity.

Our second result is about typing: there is a relationship between the
typing of a choreography and its transformation. Intuitively,
$\simplify{C}{k}$ can be typed using a composition of the types of
$C$.  The following definition formalises this composition.  We remind
the reader that a global type can always be regarded to as a standard
regular tree representation~\cite{PIERCE}. In the sequel, the function
$\paths{G}$ denotes the set of paths in the regular tree
representation of a global type $G$. Moreover, the function
$\mathsf{interleave}$ applied to a set of paths returns the set of all
their possible interleaves. ${}^*$ is the standard Kleene star,
denoting closure of paths under repetition.
%
%
\begin{definition}[Mesh Global Types]
  Given a set of global types $\{G_1,\ldots,G_n\}$, we define
  $\mesh{\{G_1,\ldots,G_n\}}$, called the {\em mesh} of
  $G_1,\ldots,G_n$, as the closure under $\alpha$-renaming of the set
  \[\textstyle \{\ G\ \pp\ p\in\mathsf{paths}(G) \ \textit{ only if }\ 
  p\in\mathsf{interleave}(p_1^*,\ldots,p_m^*)\text{ for some
  }p_j\in\bigcup_{1\leq i\leq n}\paths{G_i}\ \}\]
\end{definition}
The mesh of a set of global types is the set of all the global types whose paths are
the interleaving of some repetitions of the paths of the original types.
We can now state our second main result: the transformation of a well-typed
choreography is still well-typed and its type is in the mesh of the original types.
Below, $\tproves{a_1:G_1,\,\ldots\,,a_n:G_n}{\ C\ }{k_{n+1}:G_{n+1},\,\ldots\,,k:G_{m}}$
refers to the type system found in~\cite{CM12}. Intuitively, $C$ is well-typed if it
follows the protocols described by $G_i$ in each session to be started through $a_i$ and
each running session $k_i$.
\begin{theorem}[Transformation Typing]
  Let $C$ be a choreography such that
  \[\tproves{a_1:G_1,\,\ldots\,,a_n:G_n\quad}{\quad
C\quad}{\quad k_{n+1}:G_{n+1},\,\ldots\,,k:G_{m}}\] Then, for every session channel name
$k$ there exists $G \in \mesh{\{G_1,\,\ldots\,,G_{m}\}}$ such that
  \[\tproves{\emptyset \quad}{\quad \simplify{C}{k}\quad }{\quad k:G}\]
\end{theorem}
%
%
Considering again our example, we can type its transformation
$\simplify Ck$ with the following global type $G$.
\[
  G\ = \ 
  \REC {\ t\ } {}
  \quad \interact{\role u}{\role c}{\typestring};\ 
  \interact{\role c}{\role f}{\typestring};\ 
  \interact{\role f}{\role c}{
    \{
    \labels{ok}: \interact{\role c}{\role f}{\typefile},\quad
\labels{quit}: \RECVAR t
\}
}
\]
Observe that type $G$ is a nontrivial composition of the types $G_a$
and $G_b$ that we have shown in our introduction. Indeed we can
observe that $\role c$ is a single role even though thread $\thr c$
plays two protocols in the original choreography.

%
%

\section{Conclusions and Further Developments}
\label{sec:conclusion}
We have shown how a choreography implementing different sessions of
different types can be transformed into a choreography with a single
session implementing a single global type. Furthermore, the type of
the latter is a composition of the original types.

Our transformation is useful for eliciting the abstract behaviour of a
system that implements many protocols (given as types). A programmer
may design a choreography and then check if the global abstract
behaviour of its implementation is the expected one. Even
more importantly, a software architect could exploit our
transformation in order to design new standard protocols by extracting
them from a choreography. Interestingly, our transformation could also be used
for extracting global types out of binary session types once a global
implementation is given~\cite{CHY12}.

Another potential benefit of our transformation lies in resource control.
Our transformation removes protocol starts but preserves behaviour. This has two
implications.
First, all the synchronisations required for starting a protocol instance at the endpoint
level are no longer required. This may help in improving the performance of a system.
Second, in practice
it is usually the case that threads (or processes) can be dynamically spawned
at runtime whenever a new session is created.
We believe that our transformation can be extended to transform a choreography with an
unbounded number of threads and sessions (due to recursion) to a choreography with a
finite number of these resources. This would help in managing the resource consumption of
complex distributed systems, leading to applications, e.g., in the
fields of embedded systems and optimisation.  For example, one could
design an ad-hoc system optimised for a choreography with a predetermined
number of threads.

The formalisation presented in this paper is only an initial step towards a more complete
theory. Specifically, we did not deal with some useful features such as channel passing
and thread spawning. We plan to investigate these features in future work.
We also plan to give a concrete implementation of our transformation for the
\Chor{} language~\cite{CM12,chor:website}, and use it to benchmark our theory
through practical scenarios.

\label{sec:bib}
\bibliographystyle{eptcs}
\bibliography{biblio}

\end{document}
